\documentclass[letterpaper, 10 pt, conference]{template/ieeeconf}

\IEEEoverridecommandlockouts

\usepackage{cite}
\usepackage{amsmath,amssymb,amsfonts}
\usepackage{algorithmic}
\usepackage{graphicx}
\usepackage{textcomp}
\usepackage{xcolor}
\usepackage[nolist,nohyperlinks]{acronym}
\def\BibTeX{{\rm B\kern-.05em{\sc i\kern-.025em b}\kern-.08em
    T\kern-.1667em\lower.7ex\hbox{E}\kern-.125emX}}

\usepackage{tikz}
\usepackage{pgfplots}
\usepackage{booktabs}
\usepackage{makecell}
\usepackage{csquotes}
\pgfplotsset{compat=1.9}
\usepgfplotslibrary{statistics}
\usetikzlibrary{arrows}
\usetikzlibrary{decorations.pathreplacing,decorations.markings}
\usetikzlibrary{positioning}
\usetikzlibrary{calc}
\usetikzlibrary{fit}

\usepackage[americanvoltages]{circuitikz}
\ctikzset{bipoles/length=0.85cm}
\ctikzset{bipoles/thickness=1}

\usepgfplotslibrary{external}
\tikzset{external/system call={pdflatex \tikzexternalcheckshellescape -halt-on-error
        -interaction=batchmode -jobname "\image" "\texsource"}}
\pgfplotsset{every axis/.append style={semithick,tick style={major tick
            length=4pt,semithick,black}}}

\pgfkeys{/pgfplots/x axis shift down/.style={
        x axis line style={yshift=-#1},
        xtick style={yshift=-#1},
        tick align=outside,
        xticklabel shift={#1}}}

\pgfkeys{/pgfplots/y axis shift left/.style={
        y axis line style={xshift=-#1},
        ytick style={xshift=-#1},
        yticklabel shift={#1},
        scaled y ticks = false,
        tick align=outside,
        y tick label style={/pgf/number format/fixed,
        }
    }
}

\pgfplotsset{myPlot/.style={%
        width=8cm,
        height=3.5cm,
        line width = 0.7pt,
        separate axis lines,
        axis x line*=bottom,
        x axis shift down = 3pt,
        enlarge x limits=false,
        axis y line*=left,
        y axis shift left = 6pt,
        enlarge y limits={abs=.25pt},
        enlarge x limits={abs=.25pt},
    }
}

\definecolor{vgRed}{RGB}{193, 48, 24}
\definecolor{vgOrange}{RGB}{243, 111, 19}
\definecolor{vgYellow}{RGB}{235, 203, 56}
\definecolor{vgGreen}{RGB}{162, 185, 105}
\definecolor{vgLightBlue}{RGB}{13, 149, 188}
\definecolor{vgDarkBlue}{RGB}{6, 56, 81}

\definecolor{hks}{RGB}{199,16,92}
\definecolor{Dunkelgr}{RGB}{21, 56, 36}
\definecolor{Hellblau}{RGB}{80, 149, 200}
\definecolor{Gelbgrue}{RGB}{196, 210, 15}
\definecolor{Hellgrue}{RGB}{74, 172, 150}
\definecolor{Goldgelb}{RGB}{234, 195, 114}

\newtheorem{remark}{Remark}

\newcommand{\imag}{\ensuremath{\mathrm{\imath}}}

\usepackage{hyperref}

\title{\LARGE \bf
Identification of Power Systems with Droop-Controlled Units Using Neural Ordinary Differential Equations}

\author{Hannes M. H. Wolf$^{1}$ and Christian A. Hans$^{1}$% <-this % stops a space
\thanks{$^{1}$Hannes M. H. Wolf and Christian A. Hans are with the
    Automation and Sensorics in Networked Systems Group,
    University of Kassel,
    Germany,
    {\tt \href{mailto:h.wolf@uni-kassel.de}{h.wolf@uni-kassel.de}},
    {\tt \href{mailto:hans@uni-kassel.de}{hans@uni-kassel.de}}.}%
}

\begin{document}

\maketitle

\begin{acronym}
    \acro{ml}[ML]{machine learning}
    \acro{nn}[NN]{neural network}
    \acro{ivp}[IVP]{initial value problem}
    \acro{node}[NODE]{neural ordinary differential equation}
    \acro{pinn}[PINN]{physics informed neural network}
    \acro{pde}[NHNN]{nearly hamiltonian neural network}
    \acro{sindy}[SINDy]{sparse identification of nonlinear dynamics}
    \acro{der}[DER]{distributed energy resource}
    \acro{ho}[HO]{hyperparameter optimization}
    \acro{rmse}[RMSE]{root-mean-square-error}
    \acro{sg}[SG]{synchronous generator}
    \acro{ode}[ODE]{ordinary differential equation}
    \acro{mlp}[MLP]{multilayer perceptron}
    \acro{mse}[MSE]{mean squared error}
    \acro{smib}[SMIB]{single-machine-infinite-bus}
    \acro{nhnn}[NHNN]{nearly hamiltonian neural network}
    \acro{dp5}[DOPRI5]{fifth-order Dormand-Prince}
    \acro{rk4}[RK4]{fourth-order Runge-Kutta}
\end{acronym}

% ----------------------------------------------------------- Abstract ----------------------------------------------------------- 

\begin{abstract}  
    In future power systems, the detailed structure and dynamics may not always be fully known.  
    This is due to an increasing number of distributed energy resources, such as photovoltaic generators, battery storage systems, heat pumps and electric vehicles, as well as a shift towards active distribution grids.
    Obtaining physically-based models for simulation and control synthesis can therefore become challenging. 
    Differential equations, where the right-hand side is represented by a neural network, i.e., \acp{node}, have a great potential to serve as a data-driven black-box model to overcome this challenge. 
    This paper explores their use in identifying the dynamics of droop-controlled grid-forming units based on inputs and state measurements.
    In numerical studies, various \acp{node} structures used with different numerical solvers are trained and evaluated. 
    Moreover, they are compared to the \ac{sindy} method. 
    The results demonstrate that even though \ac{sindy} yields more accurate models, \acp{node} achieve good prediction performance without prior knowledge about the system's nonlinearities which \ac{sindy} requires to work best.
\end{abstract}

% \begin{IEEEkeywords}
% System identification, neural ordinary differential equations, droop control, power systems.
% \end{IEEEkeywords}

\acresetall
% ----------------------------------------------------------- Intorduction chapter ----------------------------------------------------------- 

\section{Introduction}
For a reliable operation of power grids, accurate dynamical models are of great importance. They enable simulations, control synthesis and prediction of future system behaviour. 
Tradionally, power systems were composed of a relatively small number of large-scale \acp{sg} connected by high- or medium-voltage transmission lines, providing power to typically passive loads. 
In this context, physical-based modeling has been a viable approach.
However, with the transition towards low carbon energy systems, this is no longer the case. 
The increasing share of small-scale distributed energy ressources, as well as the transition from consumer- to prosumer-based active distribution system, result in increasingly complex and partly uncertain dynamics. 
Unknown system parameters including inaccurate low-voltage grid impedances and dynamics of distributed actors render physically-based modeling impractical.

With the wide deployment of sensors in power systems, data-driven methods for characterizing its dynamic behaviour appear promising. 
These rely dominantly on data and do not require detailed knowledge of the underlying processes or parameters. 
Modern power systems, especially active distribution grids, are subject to changes due to additions or removals of components as well as degradation, e.g., of storage units units. Data-driven methods could deal with this by retraining models with updated data or transfer learning. 

Data-driven system identification can be partitioned into gray- and black-box approaches.
% In studies targeting data-driven identification, mainly gray-box and black-box approaches are pursued. 
While the latter solely employ data, the former leverage some physical knowledge and data to identify system dynamics. 

Gray-box approaches have gained popularity in recent years. 
\Acp{pinn}, for example, learn solutions to initial value problems based on data and a physical system description with possibly unknown parameters.
The first  application towards power systems was published in \cite{misyris_physics-informed_2020}, where the rotor angle and frequency, as well as uncertain parameters of a single-machine-infinite-bus setup were learned. 
The authors of \cite{zhang_learning_2023} propose a nearly Hamiltonian neural network, implicitely embedding energy conservation laws in the network architecture. 
By learning the Hamiltonian of the system and using automatic differentiation, the frequency dynamics of a single-machine-infinite-bus steup are identified with focus on fault scenarios. 
Moreover, \ac{sindy} has been used in \cite{nandakumar_sparse_2022} to learn the frequency and voltage dynamics for a microgrid under disturbances, e.g., load variations, based on state measurements. 
The authors of \cite{saeed_kandezy_sindy_2024} use voltage measurements to learn the dynamics of a power grid for load variations and fault scenarios with \ac{sindy}.  
Lastly, the authors of \cite{buttner_complex-phase_2024} use a gray-box model for grid-forming units, called normal form, to identify the dynamics of a single grid forming inverter under fault conditions. 

Black-box approaches, on the other hand, rely solely on data. 
The proposed method in \cite{zhao_structure-informed_2022}, for example, does not require physical knowledge about units or the grid, but makes use of the fact, that power systems follow a graph structure to predict post-fault trajectories online using only historic data. 
Similarily, the authors of \cite{sharma_data-driven_2019} predict state trajectories based on historic data by making use of the Koopman operator.
Since the introduction of \acp{node} in \cite{chen_neural_2018}, they have been used in multiple fields and applications. 
One of them is black-box system identification which has been conducted, e.g., in \cite{rahman_neural_2022} where the authors use them to model the input-output dynamics of different systems.
In the power system domain, \acp{node} have also been employed for example in \cite{xiao_feasibility_2023} and \cite{zhang_learning_2024}. 
In \cite{xiao_feasibility_2023}, models are learned from portal measurements in order to create dynamic equivalents of the power system components which enable integrated transient simulation. 
In \cite{zhang_learning_2024}, post-fault frequency dynamics of grid-connected \acp{sg} are identified.

It is important to note, that \acp{pinn}, as used in \cite{misyris_physics-informed_2020}, learn solutions to initial value problems which includes the estimation of unknown parameters. 
Resulting solutions account for specific initial conditions and inputs. 
Consequently, the learned \ac{nn} can not be used for simulations in different settings \cite{legaard_constructing_2022}. 
Furthermore, they require detailed knowledge about the underlying differential equations in order to guide the learning process. 
Similarly, \ac{sindy}, as used in \cite{nandakumar_sparse_2022} and \cite{saeed_kandezy_sindy_2024}, requires good guesses about the nonlinear terms occuring in the system's differential equations in order to work well. 
Another limitation of existing approaches, e.g. \cite{xiao_feasibility_2023} and \cite{buttner_complex-phase_2024}, is that only single components or dynamic equivalents, but not entire power system dynamics are identified. 
While purely data-driven approaches are successfully employed in \cite{zhao_structure-informed_2022,sharma_data-driven_2019}, they learn models for time-series prediction based on historic state measurements, rather then identifying continuous-time system dynamics. 
Finally, the approaches in \cite{zhang_learning_2024} and \cite{zhang_learning_2023} are limited to trajectory predictions in fault scenarios and and do not consider control inputs. 
This results in models that are not suitable for control synthesis and closed-loop simulation.
Lastly, most publications provide little insights into what structures work well and what more general conclusions can be drawn. 
To our knowledge, only \cite{xiao_feasibility_2023} conducts a limited, gridsearch-like method to find appropriate \ac{nn} structures, but does not conduct an extensive hyperparameter optimization.

The contributions of this paper are as follows:
1) We apply \acp{node} to identify the dynamics of an entire power system containing droop-controlled grid-forming units using input and state trajectories. This includes dynamics of the nodes, as well as their nonlinear coupling through the electrical network.
    We learn the dynamics from nodal measurements, enabling a prediction of voltage and frequency dynamics of the system. 
2) While comparable studies focus on predicting trajectories of autonomous dynamical systems under fault scenarios, we aim to learn models that can be used for control synthesis and closed-loop simulation, i.e, input-output systems.
3) We draw a comparison between the performance of \acp{node} and \ac{sindy} for the system at hand and show that, even though \ac{sindy} is more accurate, good prediction accuracy can be achieved with \acp{node} without relying on physical knowledge about the system. 
4) We conduct an extensive hyperparameter optimization to investigate the influence of the \ac{nn} size and the activation functions on the accuracy of the models. We conclude that smaller networks with a continuous activation function are advantageous for the task at hand.
5) We investigate the use of different integration schemes for \acp{node} and show that using higher-order solvers does not necessarily result in more accurate predictions.

The remainder of this paper is structured as follows. 
First, the power system model is presented. 
Then, the identification methods, i.e., \acp{node} and \ac{sindy}, are introduced. 
Consecutively, we describe the numerical experiments and conclude the paper with a discussion of the results and an outlook on future work.

% ----------------------------------------------------------- Modeling Chapter  ----------------------------------------------------------- 
\section{Modeling}
\label{sec:modeling}
In this section, the power system model of the droop-controlled grid-forming units is described. 
First, the power flow equations of the network, that couples the units and loads, are presented. 
Then, the dynamics of the grid-forming units are described. 
Finally, a state transformation for the voltage angles is performed and the overall nonlinear state model of the coupled grid-forming units is posed. 

\subsection{Notation}
We make use of the following notation: $\mathbb{N}$ denotes the set of positive integers, $\mathbb{R}$ denotes the set of real numbers and $\mathbb{R}_{\geq 0}$ denotes the set of positive real numbers.
Moreover, $\mathbb{C}$ denotes the set of complex numbers with $\imag$ being the imaginary unit. The cardinality of a set $\mathbb{V}$ is denoted by $|\mathbb{V}|$.
Moreover, $\| \cdot \|_2$ refers to the Euclidean norm.

\subsection{Network}
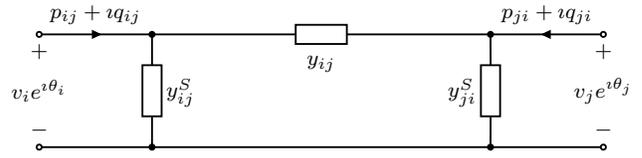
\begin{figure}
    \centering
    % !TeX encoding = UTF-8
% !TeX spellcheck = en_GB
% !TEX TS-program = pdflatex
% !TEX root = ../document.tex

\tikzset{external/export next=true}
\begin{tikzpicture}[xscale=1.5, yscale = 1.0, font=\footnotesize, line width=0.7pt]
		\draw
		(0, 0) to[open, v^<={$v_i e^{\imag\theta_i}$}] (0, 1.5)
		(0, 0) to[short, o-] (1, 0) to[short, -] (4, 0) to[short, -o] (5, 0)
		(0, 1.5) to[short, o-, i^>={$p_{ij} + \imag q_{ij}$}] (1, 1.5)
		(1, 0) to[generic, l_=$y_{ij}^S$, *-*] (1, 1.5)
		(1, 1.5) to[short] (2.2, 1.5) to[generic, l_=$y_{ij}$] (2.8, 1.5) to[short] (4, 1.5)
		(4, 0) to[generic, l^=$y_{ji}^S$, *-*] (4, 1.5)
		(5, 1.5) to[short, o-, i_>={$p_{ji} + \imag q_{ji}$}] (4, 1.5)
		(5, 0) to[open, v<={$v_j e^{\imag\theta_j}$}] (5, 1.5)
		;

\end{tikzpicture}
    \caption{Power, voltage and impedances between two grid-forming nodes.}
    \label{fig:twoGridFormingNodesPi}
\end{figure}
Consider a connected graph $\mathbb{G} = (\mathbb{V}, \mathbb{E})$, where $\mathbb{V} \subset \mathbb{N}$ is the set of nodes and $\mathbb{E} \subseteq \left[\mathbb{V}\right]^2$ the set of edges. 
The set of nodes adjacent to $i \in \mathbb{V}$ is referred to as $\mathbb{V}_i \subset \mathbb{V}$. 
We assume Kron-reduced grids \cite{dorfler_kron_2013}, where each node $i$ is associated with a grid-forming unit and each edge $e=(i,j) \in \mathbb{E}$ with a transmission line. 
The $\pi$-equivalent circuit of such a transmission line from node $i$ to node $j \in \mathbb{V}$ is shown in Fig.~\ref{fig:twoGridFormingNodesPi}. 
It consists of the parallel shunt admittance $y_{ij}^S = g_{ij}^S + \imath b_{ij}^S \in \mathbb{C}$ and the series impedance $y_{ij} = g_{ij} + \imath b_{ij} \in \mathbb{C}$, where $g \in \mathbb{R}_{\geq 0}$ and $b\in \mathbb{R}$ denote the conductance and susceptance, respectively.
The active and reactive power flow, $p_{ij}\in \mathbb{R}$ and $q_{ij} \in \mathbb{R}$, from node $i$ to node $j$ over such a transmission line is \cite{kundur_power_1994}
\begin{subequations}
    \label{eq:power_flow}
    \begin{align}
        p_{ij} &= v_i^2 \, (g_{ij} + g_{ij}^S) - v_i v_j\left(g_{ij} \cos (\delta_{ij}) + b_{ij} \sin (\delta_{ij})\right), \\
        q_{ij} &= - v_i^2 \, (b_{ij} + b_{ij}^S) - v_i v_j\left(g_{ij} \sin (\delta_{ij}) - b_{ij} \cos (\delta_{ij})\right).
    \end{align}
\end{subequations}
Here, $v_i \in \mathbb{R}$ denotes the voltage amplitude and $\delta_{ij} = \delta_i - \delta_j$ the difference between the voltage phase angles $\delta_i \in \mathbb{R}_{\geq 0}$ and $\delta_j \in \mathbb{R}_{\geq 0}$ at nodes $i$ and $j$.
Furthermore, we consider constant impedance loads $y_{i}^L = g_{i}^L + \imath b_{i}^L \in \mathbb{C}$ at node $i$.
The overall active and reactive power at node $i$ is \cite{schiffer_conditions_2014}
\begin{subequations}
    \label{eq:overall_power}
    \begin{align}
        p_{i} &= v_i^2 \, g_{i}^L + \textstyle\sum\limits_{j \in \mathbb{V}_i} p_{ij}, \\
        q_{i} &= -v_i^2 \, b_{i}^L+ \textstyle\sum\limits_{j \in \mathbb{V}_i} q_{ij}. 
    \end{align}
\end{subequations}

\subsection{Droop-controlled grid-forming inverters}
Following standard practice \cite{schiffer_conditions_2014}, we assume that the voltage amplitude and frequency can be set instantaneously. 
Applying proportional droop-control yields equations of the form 
\begin{subequations}
    \label{eq:droop_control_dynamics}
    \begin{align}
        \dot{\delta}_i&= \omega_i^d - k_i^P (p_i^m - p_i^d), \\
        v_i  &= v_i^d - k_i^Q (q_i^m - q_i^d),
    \end{align}
where $\omega_i^d \in \mathbb{R}_{\geq 0}$ and $v_i^d \in \mathbb{R}_{\geq 0}$ denote the desired frequency and voltage amplitude at node $i$, respectively. 
The desired active and reactive power, denoted by $p_i^d \in \mathbb{R}$ and $q_i^d \in \mathbb{R}$, are typically provided by a high-level tertiary control or energy management \cite{hans_operation_2021}.
The constants $k_{i}^P \in \mathbb{R}_{\geq 0}$ and $k_{i}^Q \in \mathbb{R}_{\geq 0}$ denote the droop gains. 
We further assume, that active and reactive power measurements are associated with first-order filters of the form \cite{schiffer_conditions_2014}
   \begin{align}
        \tau_{i}^P \dot{p}_i^m &= - p_i^m + p_i, \\
        \tau_{i}^Q \dot{q}_i^m &= - q_i^m + q_i, 
   \end{align} 
\end{subequations}
where $p_i^m \in \mathbb{R}$ and $q_i^m \in \mathbb{R}$ denote the measured active and reactive power, and $\tau_{i}^P \in \mathbb{R}_{\geq 0}$ and $\tau_{i}^Q \in \mathbb{R}_{\geq 0}$ are filter time constants.

As discussed in \cite{schiffer_synchronization_2013} \cite{schiffer_conditions_2014}, droop-controlled \acp{sg} can also be modeled in a form reminiscent of~(\ref{eq:droop_control_dynamics}). 
Thus, the dynamics~(\ref{eq:droop_control_dynamics}) can represent droop-controlled inverters or \acp{sg}.

\subsection{Change of states to voltage angles differences}
\label{sec:change_of_states}
Similar to applications where a slack bus is used, we take the phase angle of one node as reference and describe the remaining angles as differences.
This is a reasonable approach, since the phase angle differences and not the actual phase angles cause changes in the powerflow~(\ref*{eq:power_flow}).  
We choose $\delta_1$, as the reference and drop its state from the system dynamics. 
All other phase angles are described by phase angle differences $\delta_{1i}$. 
This allows us to reduce the dimension of the system by one. 
For node $i \in \mathbb{V} \setminus \{1\}$, this yields 
\begin{align}
    \label{eq:change_of_states}
    \dot{\delta}_{1i} &= \omega_{1}^d - k_1^P (p_1^m - p_1^d) - \omega_{i}^d + k_i^P (p_i^m - p_i^d)
\end{align}    

\subsection{Overall dynamics}

Combining (\ref*{eq:power_flow}), (\ref*{eq:overall_power}), (\ref*{eq:droop_control_dynamics}) and (\ref{eq:change_of_states}), we obtain the overall dynamics of the coupled grid-forming units in form of a set of \acp{ode}. 
For all nodes $i \in \mathbb{V}$, the dynamics are given by
\begin{subequations}
    \label{eq:overall_dynamics}
    \begin{align}
        \dot{\delta}_{1i} &= \omega_{1}^d - k_1^P (p_1^m - p_1^d) - \omega_{i}^d + k_i^P (p_i^m - p_i^d), \label{eq:overall_theta} \\
        \dot{p}_i^m &= \textstyle\dfrac{1}{\tau_{i}^P} (- p_i^m + p_{ii} + \textstyle\sum\limits_{j \in \mathbb{V}_i} p_{ij}), \\
        \dot{v}_i &= \textstyle\dfrac{1}{\tau_i^Q} ( -v_i + v_i^d - k_i^Q (q_{ii} + \textstyle\sum\limits_{j \in \mathbb{V}_i} q_{ij} - q_i^d )) .
    \end{align}
\end{subequations}
\begin{remark}
    For node $i=1$, (\ref*{eq:overall_theta}) becomes zero and is thus, according to the descriptions in Sec.~\ref{sec:change_of_states}, omitted overall model.
\end{remark}

For what follows, it is convenient to describe~(\ref{eq:overall_dynamics}) in the compact form
\begin{equation}
    \label{eq:nonlinear_state_model}
    \dot{x}(t) = f(x(t),u(t))
\end{equation}
with state $x \in \mathbb{R}^{N_x}$, $N_x=3|\mathbb{V}|-1$ and control input $u  \in \mathbb{R}^{N_u}$, $N_u=2|\mathbb{V}|$.  
In detail, the state and control input are
\begin{align*}
    x &= \begin{bmatrix} \delta_{12} & \ldots & \delta_{1N} & p_1^m & \ldots & p_N^m & v_1 & \ldots & v_N \end{bmatrix}^T, \\
    u &= \begin{bmatrix} v_1^d & \ldots & v_N^d & \omega_1^d & \ldots & \omega_N^d \end{bmatrix}^T.
\end{align*}
The remaining parts of (\ref*{eq:overall_dynamics}), which are not elements of $x$ or $u$, are assumed to be constant parameters of the system.

\section{Methods}
We we aim to learn dynamics of the form~(\ref*{eq:nonlinear_state_model}) using data-driven system identification methods. 
In detail, we will use \acp{node}, a black-box method, as well as \ac{sindy}, which is a state-of-the-art gray-box method for nonlinear systems. 
In what follows, both will be recalled.

\subsection{Neural ordinary differential equations (NODEs)}
\acp{node} were first proposed in \cite{chen_neural_2018}. 
A \ac{node} is a differential equation, where the vector field on the right hand side of the continuous-time dynamics 
\begin{equation}
    \label{eq:node}
    \dot{x}(t) =  \mathcal{N}_\Theta(t, x(t))
\end{equation}
is represented by a \ac{nn} $\mathcal{N}_\Theta$  with parameters $\Theta$.
Typically, standard structures like feedforward \acp{nn} are used for \acp{node}~\cite{kidger_neural_2022}. 

\subsubsection{Solving the initial value problem}
\begin{figure}
    \vspace{0.6em}
    \centering
    % !TeX spellcheck = en_US
% !TeX encoding = UTF-8
% !TEX root = ../document.tex

\tikzset{external/export next=true}

\tikzset{blockScheme/.style={%
  >=latex,
  line width = 0.65pt,
  color=black,
  node distance = 10mm,
  }
}

\tikzset{square/.style = {draw, minimum height=1.5em, minimum width=2.5em, outer sep=0pt, node distance=0em, inner sep=3pt}}

\tikzset{block/.style={draw, minimum height=2.0em, minimum width=2.0em, fill=white, outer sep=0pt, inner sep=3pt, align=center}}

\tikzset{blockScheme timeStepper/.style={blockScheme, node distance = 1.7cm, line width=0.5pt}}

\newcommand{\inputnum}{3} 
\newcommand{\hiddennum}{3}  
\newcommand{\outputnum}{2}

\begin{tikzpicture}[blockScheme timeStepper, font=\footnotesize]

\node[square] (xu0) {$x(t_s)$};
\node[square, right=of xu0] (xu1) {${x}(\tilde{t}_1)$};
\node[square, right=of xu1] (xu2) {${x}(\tilde{t}_2)$};
\node[square, right=of xu2] (xu3) {${x}(\tilde{t}_3)$};
\node[square, right=of xu3] (xudot) {$\ldots$};
\node[square, right=of xudot] (xuK) {${x}(t_e)$};

\node[block, node distance=6mm,  below = of xu1, xshift = 10mm] (int) {ODE solver};
\draw[->] (xu0) |- (int);

\node[block, node distance=5mm, right = of int] (nn) {$\mathcal{N}_\Theta$};
\draw[->] ($(nn.west)+(0, 0.15)$) to ($(int.east)+(0, 0.15)$);
\draw[<-] ($(nn.west)+(0, -0.15)$) to ($(int.east)+(0, -0.15)$);

\node[anchor=north] at (2.68, -0.15) (brace) {$\underbrace{\hspace{44mm}}$};

\draw[->, bend right=5] (int) to (2.65, -0.5);

\node[fill = none, draw, dashed, fit=(xu0)(xuK)(int)] (outline) {};

\draw[<-] (outline.west) to node[above] {$x(t_s)$} ++(-1, 0);

\draw[->] (outline.east) to node[above] {$x(t_e)$} ++(1, 0);

\end{tikzpicture}
    \caption{Solving the \ac{ivp} with initial condition $x(t_s)$ and $\mathcal{N}_\Theta$ for $t_e$. Here, $\tilde{t}_1 < \tilde{t}_2 < \tilde{t}_3 < \dots$ denote the internal evaluation times chosen by the solver. Illustration motivated by~\cite{legaard_constructing_2022}.}
    \label{fig:solve_ivp}
\end{figure}
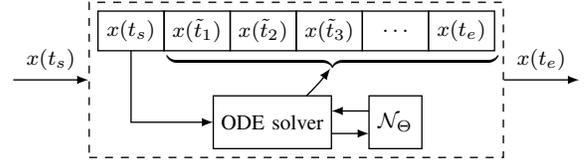
\begin{figure}
    \centering
    % !TeX spellcheck = en_US
% !TeX encoding = UTF-8
% !TEX root = ../document.tex

\tikzset{external/export next=true}

\tikzset{blockScheme/.style={%
  >=latex,
  line width = 0.65pt,
  color=black,
  node distance = 10mm,
  }
}

\tikzset{square/.style = {draw, minimum height=2.5em, minimum width=2.5em, outer sep=0pt, node distance=0em, inner sep=3pt}}

\tikzset{block/.style={draw, minimum height=2.0em, minimum width=2.0em, fill=white, outer sep=0pt, inner sep=0pt, align=center}}

\tikzset{blockScheme timeStepper/.style={blockScheme, node distance = 1.7cm, line width=0.5pt}}

\newcommand{\inputnum}{3} 
\newcommand{\hiddennum}{3}  
\newcommand{\outputnum}{2}

\begin{tikzpicture}[blockScheme timeStepper, font=\footnotesize]

\node[square] (xu0) {$\begin{bmatrix}{x}(t_0) \\ u(t_0) \end{bmatrix}$};
\node[square, right=of xu0] (xu1) {$\begin{bmatrix}{x}(t_1) \\ u(t_1) \end{bmatrix}$};
\node[square, right=of xu1] (xudots) {$\ldots$};
\node[square, right=of xudots] (xuKm1) {$\begin{bmatrix}{x}(t_{K-1}) \\ u(t_{K-1}) \end{bmatrix}$};
\node[square, right=of xuKm1] (xuK) {${x}(t_K)$};

\node[block, node distance=3mm, below = of xu1.south west] (ivp0) {solve \\ IVP};
\draw[->] (xu0) |- (ivp0);
\draw[<-] (xu1.south)++(-0.1, 0) |- (ivp0);

\node[block, node distance=3mm,  below = of xudots, draw=none, fill=none, minimum width=3.5em] (ivpdots) {$\ldots$};
\draw[->] (xu1.south)++(0.1, 0) |- (ivpdots);
\draw[<-] (xuKm1.south)++(-0.1, 0) |- (ivpdots);

\node[block, node distance=3mm, below = of xuK.south west] (ivpKm1) {solve \\ IVP};
\draw[->] (xuKm1.south)++(0.1, 0) |- (ivpKm1);
\draw[->] (ivpKm1) -| (xuK);

\node[block, node distance=1mm, below = of ivpdots] (nn) {$\mathcal{N}_\Theta$};

\node[fill = none, draw, dashed, fit=(xu0)(xuK)(nn)] (outline) {};

\draw[<-] (outline.north west)++(0, -0.4) to node[below, xshift=-0.7em, yshift=-0.3em, inner sep=0pt,] {$\begin{bmatrix}x(t_0) \\ u(t_0) \\ u(t_1) \\ \vdots \\ u(t_{K-1})\end{bmatrix}$} ++(-1, 0);

\draw[->] (outline.north east)++(0, -0.4) to node[below, xshift=0.1em, yshift=-0.3em, inner sep=0pt,] {$\begin{bmatrix}x(t_0) \\ {x}(t_1) \\ {x}(t_2) \\ \vdots \\ {x}(t_K)\end{bmatrix}$} ++(1, 0);

\draw[<->, bend left=5] (nn) to (ivp0);
\draw[<->, bend right=5] (nn) to (ivpKm1);
\draw[<->, bend right=5] (ivpdots.west)++(0, -0.1) to (nn);
\draw[<->, bend left=5] (ivpdots.east)++(0, -0.1) to (nn);
\draw[<->] (ivpdots.center)++(0, -0.1) to (nn);

 \begin{scope}[yshift=-20mm, xshift = 1mm, xscale=1.1, font=\scriptsize]

    \node[draw, densely dotted, fill=white, minimum height=30mm, minimum width=45mm] (nnDetail) at (1.5, -2) {};

    \draw[densely dotted] (nnDetail.south west) to (nn.south west);
    \draw[densely dotted] (nnDetail.south east) to (nn.south east);
    \draw[densely dotted] (nnDetail.north west) to (nn.north west);
    \draw[densely dotted] (nnDetail.north east) to (nn.north east);

   \node[draw, densely dotted, fill=white, minimum height=30mm, minimum width=45mm] (nnDetail) at (1.5, -2) {};

    \node[circle, inner sep=0pt, minimum size = 6mm, fill=Gelbgrue!50] (Input-1) at (0,-1) {$x_1$};
    \node[circle, inner sep=0pt, minimum size = 6mm, fill=Gelbgrue!50] (Input-2) at (0,-2) {$x_2$};
    \node[circle, inner sep=0pt, minimum size = 6mm, fill=Gelbgrue!50] (Input-3) at (0,-3) {$u$};
     
    % Hidden Layer
    \foreach \i in {1,...,\hiddennum}
    {
        \node[circle, 
            minimum size = 6mm,
            fill=Hellblau!50,
            yshift=(\hiddennum-\inputnum)*5 mm
        ] (HiddenOne-\i) at (1,-\i) {};
    }

    % Hidden Layer
    \foreach \i in {1,...,\hiddennum}
    {
        \node[circle, 
            minimum size = 6mm,
            fill=Hellblau!50,
            yshift=(\hiddennum-\inputnum)*5 mm
        ] (HiddenTwo-\i) at (2,-\i) {};
    }

    % Output Layer
    \node[circle, 
        minimum size = 6mm,
        fill=Hellgrue!50,
        inner sep=0pt,
        yshift=(\outputnum-\inputnum)*5 mm
    ] (Output-1) at (3,-1) {$\frac{\partial x_1}{\partial t}$};

    \node[circle, 
        minimum size = 6mm,
        fill=Hellgrue!50,
        inner sep=0pt,
        yshift=(\outputnum-\inputnum)*5 mm
    ] (Output-2) at (3,-2) {$\frac{\partial x_2}{\partial t}$};

    % Connect neurons In-Hidden
    \foreach \i in {1,...,\inputnum}
    {
        \foreach \j in {1,...,\hiddennum}
        {
            \draw[->, shorten >=0.1pt] (Input-\i) -- (HiddenOne-\j);   
        }
    }
     
    % Connect neurons In-Hidden
    \foreach \i in {1,...,\hiddennum}
    {
        \foreach \j in {1,...,\hiddennum}
        {
            \draw[->, shorten >=0.1pt] (HiddenOne-\i) -- (HiddenTwo-\j);   
        }
    }

    % Connect neurons Hidden-Out
    \foreach \i in {1,...,\hiddennum}
    {
        \foreach \j in {1,...,\outputnum}
        {
            \draw[->, shorten >=0.1pt] (HiddenTwo-\i) -- (Output-\j);
        }
    }

\end{scope}

\end{tikzpicture}
    \caption{Repeatedly solving an \ac{ivp} with different initial conditions and inputs to step the model forward in time. Illustration motivated by~\cite{legaard_constructing_2022}. Note that the portrayed \ac{nn} serves as an example and does not show the correct dimensions of the actual model.}
    \label{fig:neural_odes}
\end{figure}
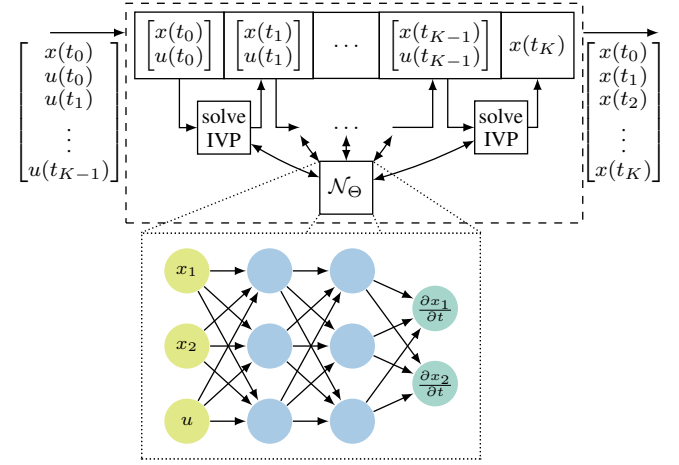
Taking~\eqref{eq:node} together with an initial value $x(t_s)$ at time $t_s \in \mathbb{R}_{\geq 0}$, we obtain an \ac{ivp}. Here, $\mathcal{N}_\Theta$ can be used together with a numerical solver to find a solution $x(t_e)$, $t_s < t_e$, to this \ac{ivp} of the form \cite{chen_neural_2018}
\begin{subequations}
    \label{eq:solve_ivp}
    \begin{align}
        x(t_e) &= x(t_s) + \int_{t_s}^{t_e} \mathcal{N}_\Theta(t, x(t)) \, dt \\
        &= \texttt{ODEsolve} \left(\mathcal{N}_\Theta, (t_s,t_e), x(t_s) \right).
    \end{align}
\end{subequations}
For this task, off-the-shelf ODE solvers can be used~\cite{rahman_neural_2022} that work with explicit or implicit methods and fixed or variable step sizes. 
Fig.~\ref{fig:solve_ivp} shows the interaction of the \ac{nn} with the ODE solver.

\subsubsection{Parameterized \acsp{node}}
Relating to~(\ref{eq:nonlinear_state_model}), we seek a model that includes control inputs $u$. 
We therefore make use of the so-called parameterized \acp{node} presented in \cite{lee_parameterized_2021} and consider equidistant time intervals $[t_k, t_{k+1}]$ with $t_k=k\Delta t, \, k \in \mathbb{N}$, in which the control inputs $u(t_k)$ remain constant. 
This allows us to solve an \ac{ivp} with the initial condition $x(t_s) = x(t_k)$ from $t_s = t_k$ to $t_e = t_{k+1}$, along the lines of (\ref*{eq:solve_ivp}), by treating the control inputs as a parameters.
In practice, this is achieved by adding inputs to the \ac{nn}~\cite{lee_parameterized_2021}, that represent the control inputs, and thereby learn \enquote{a family of vector fields instead of a single one}~\cite{massaroli_dissecting_2020}. Thus, \eqref{eq:node} becomes 
\begin{equation}
    \label{eq:parametrized_node}
    \dot{x} =  \mathcal{N}_\Theta(t, x(t),u(t_k)).
\end{equation}
Fig.~\ref*{fig:neural_odes} illustrates how a model is stepped through time. For fixed-step solvers, we here set the step size to $\Delta t$.
\begin{remark}
    Using the Euler integration method, the model can be stepped forward in time with
    \begin{equation}
        x(t_{k+1}) = x(t_k) + \Delta t \, \mathcal{N}_\Theta(t_k, x(t_k), u(t_k)).
    \end{equation}
    which is referred to as an \enquote{Euler time stepper}~\cite{legaard_constructing_2022}.
\end{remark}

\subsubsection{Learning a model}
We use a \ac{mlp} to learn the vector field of the continuous-time dynamics in~(\ref*{eq:overall_dynamics}). 
The \ac{mlp} takes $x$ and $u$ as inputs and outputs state derivatives $\dot x$. 
It consists of $H$ hidden layers with $L$ hidden neurons in each layer. 
After each hidden layer, an activation function $h(\cdot)$ is applied. 
This results in the nested structure
\begin{multline}
    \label{eq:mlp}
    \mathcal{N}_\Theta(x,u) = \\ W_{o} h\left(W_H \dots h\left(W_1  [x, u]^T + b_1\right) \dots + b_H\right)+ b_{o},
\end{multline}
where $W_o \in \mathbb{R}^{N_x \times L}$, $b_o \in \mathbb{R}^{N_x}$, $W_1 \in \mathbb{R}^{L \times (N_x + N_u)}$, $W_i \in \mathbb{R}^{L \times L}$ for $i \in [2,H] \subset \mathbb{N}$ and $b_i \in \mathbb{R}^{L}$ for $i \in [1,H] \subset \mathbb{N}$ are the weights and biases, i.e., the parameters $\Theta$, of the \ac{nn}.
\begin{remark}
    Even though, \eqref{eq:node} and~\eqref{eq:parametrized_node} explicitely express a dependency of time $t$, we omit it in~\eqref{eq:mlp} as we consider a time-invariant model~\eqref{eq:overall_dynamics}.
\end{remark}
\subsubsection{Training}
To find appropriate parameters, we optimize over the residuals of the one-step-ahead prediction of the states assuming that the true state of the previous step is known. 
As the objective function, we use the \ac{mse}
\begin{multline}
    \label{eq:loss_node}
    \mathcal{L}_{ \text{NODE}} = \frac{1}{K} \sum_{k=0}^{K-1} \big\| x(t_{k+1}) - x(t_k) \big. \\ \big. - \int\displaylimits_{\hspace{-1em}t_k\hspace{-1em}}^{\hspace{-1em}t_{k+1}\hspace{-1em}} \mathcal{N}_\Theta(t, x(t),u(t_k)) dt \big\|_2^2
\end{multline}
between the true state $x(t_{k+1})$ of the reference system and the solution obtained by solving an \ac{ivp} with~\eqref{eq:mlp} using the true state $x(t_k)$ of the reference system as the initial condition. 
We train the \acp{nn} with the adjoint method which computes the gradients by solving an augmented \ac{ode} backwards in time~\cite{chen_neural_2018}. 

\subsection{System Identification of Nonlinear Dynamics (SINDy)}
\ac{sindy} has first been presented in \cite{brunton_discovering_2016} and has become popular for identification of nonlinear systems. 
\ac{sindy} is a data-driven gray-box approach which works best when including knowledge about the system.
Specifically, a set of candidate functions is used in a sparse regression problem of the form
\begin{equation}
    \dot{X} = \Xi(X,U) \Theta,
\end{equation}
with
\begin{equation}
    X = \begin{bmatrix} x^T(t_1) \\ x^T(t_2) \\ \vdots \\ x^T(t_K) \end{bmatrix}, \,   \dot{X} = \begin{bmatrix} \dot{x}^T(t_1) \\ \dot{x}^T(t_2) \\ \vdots \\ \dot{x}^T(t_K) \end{bmatrix}, \, U = \begin{bmatrix} u^T(t_1) \\ u^T(t_2) \\ \vdots \\ u^T(t_K) \end{bmatrix},
\end{equation}
where $X, \, \dot{X} \in \mathbb{R}^{K \times N_x}$  and $U \in \mathbb{R}^{K \times N_u}$ real valued matrices.
The columns of $\Xi(X,U) \in \mathbb{R}^{K \times M}$ are formed from $M \in \mathbb{N}$ candidate functions in $x$ and $u$ which are chosen, for example, based on the right-hand-side of \eqref{eq:nonlinear_state_model}. 
Moreover, $\Theta \in \mathbb{R}^{M\times N_x}$, contains the parameters that need to be learned.
The dynamics of the system at time $t$ are then approximated by a linear combination of the candidate functions weighted by elements in $\Theta$, i.e.,
\begin{equation}
    \label{eq:sindy_dynamics}
    \dot{x}(t) = \Xi(x^T(t),u^T(t)) \Theta.
\end{equation}
Similar to \acp{node}, we end up with a surrogate for the continuous-time dynamics of the system. 
Once the parameters have been learned, we can solve \acp{ivp} to obtain the solutions for the system. 
For this, we again use an off-the-shelf solver and require it to evaluate the state at similar time points as the \acp{node}.

\subsubsection{Basis functions}
For \ac{sindy} to work well, we need knowledge about the dynamics of the system. 
Specifically, we assume that we know the form of all linear and nonlinear terms that occur in~\eqref{eq:overall_dynamics}.
This comprises also the couplings in~(\ref{eq:power_flow}) for all possible powerlines between all nodes.
Specifically, we choose 
\begin{itemize}
    \item $1$ for constant terms,
    \item $\omega^d_i$, $v_i^d$, $p^m_i$, $v_i$ and $v_i^2$  for all $i \in \mathbb{V}$,
    \item $v_i v_j \sin (\delta_{ij})$ and $v_i v_j \cos (\delta_{ij})$ for all $i \in \mathbb{V}$, $j \in \mathbb{V}$ with $i < j$,
\end{itemize}
as candidate functions, which, for the case of $|\mathbb{V}|$ nodes yields a total number of $5|\mathbb{V}| + |\mathbb{V}|(|\mathbb{V}|-1) + 1$ candidate functions. 

\subsubsection{Training}
Contrarily to \acp{node}, with \ac{sindy} the model is trained by minimizing the residuals of the state derivatives. 
Therefore, a numerical differentiation method is required. 
In addition to optimizing over the residuals of the state derivatives, we enforce sparsity with a regularization of the parameters. 
For this, we choose ridge-regularization. 
The optimization can be decomposed into subproblems for each state $i \in [1, N_x]$: Let $\theta_i$ be the $i$-th column of $\Theta$ and $\dot{X}_i$ the $i$-th column of $\dot{X}$. Then the objective of each subproblem $i$ reads
\begin{equation}
    \label{eq:loss_sindy}
    \mathcal{L}_{\text{SINDy},i}  = \|\dot{X}_i - \Xi(X,U) \theta_i\|_2 + \lambda \|\theta_i\|_2^2.
\end{equation}
where $\lambda \in \mathbb{R}_{\geq 0}$ is the regularization parameter weighting the accuracy of the prediction against the sparsity of the coefficents.
Note that different to NODEs, we do not need to solve an \ac{ivp} during the training.

\section{Experiment}
\label{sec:experiment}
We conduct numerical studies to compare \acp{node} with different integration schmemes with each other and with \ac{sindy}. 
In what follows, first the system under investigation is described. 
Then, the employed data and the setup for hyperparameter optimization, as well as the training procedure are discussed.

\subsection{System under investigation}
\begin{figure}
    \vspace{0.6em}
    % !TeX encoding = UTF-8
% !TeX spellcheck = de_DE
% !TEX root = ../document.tex

\tikzset{external/export next=true}

\tikzset{mgScheme/.style={%
  >=latex,
  line width = 0.65pt,
  color=black,
  }
}

\tikzset{invS/.style={
  rectangle,
  fill=white,
  draw=black,
  inner sep=0.1pt
  }
}

\tikzset{midArrow/.style={
  draw,
  decoration={
    markings,
    mark=at position 0.5 with \arrow{stealth},%
  },
  postaction=decorate
  }
}

\input{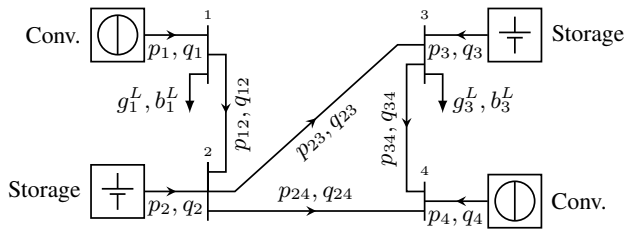}

\newcommand{\wInPlot}[1]{\ensuremath{g_{#1}^L, b_{#1}^L}}
\newcommand{\pInPlot}[1]{\ensuremath{p_{#1}, q_{#1}}}

\begin{tikzpicture}[
	mgScheme,
	yscale=1.3,
	xscale=1.2,
	font = \small,
	powerLine/.style={draw},
]

% Bar 1
\path (-2.2, 0.5) node[invS] (1) {\GraphTristateT} node[at = (1.west), left]{Conv.};

\path[midArrow] (1) -- node[below]{\pInPlot{1}} ++(1.0, 0) coordinate (bar1);
\path[powerLine] (bar1)++(0, 0.1) node[yshift = 4] {\tiny $1$}  -- ++(0, -0.6) coordinate (1r);

\path[draw, ->] (1r)++(0, 0.1) -- ++(-0.2, 0.0) -- ++(0, -0.4)  node[left, yshift=5]{\wInPlot{1}};

% Bar 2
\path (-2.2, -1.1) coordinate (2);
\path (2) ++ (1, 0) coordinate (bar2);
\path (2) ++ (0, 0) node[invS] (2s) {\GraphTristateS} node[at = (2s.west), left]{Storage};

\path[midArrow] (2s) -- node[below]{\pInPlot{2}} ++(1.0, 0) coordinate (bar2s);
\path (2) ++(1.0, 0) coordinate (bar2r);

\path[powerLine] (bar2s) -- ++(0, 0.3) node[yshift = 4] {\tiny $2$} -- (bar2r) -- ++(0, -0.3);

% Bar 3
\path (2.2, -1.2) node[invS] (3) {\GraphTristateT} node[at = (3.east), right]{Conv.};

\path[midArrow] (3) -- node[below]{\pInPlot{4}} ++(-1, 0) coordinate (bar3);
\path[powerLine] (bar3)++(0, 0.2)node[yshift = 4] {\tiny $4$} -- ++(0, -0.4);

% Bar 4
\path (2.2, 0.1) coordinate (4);
\path (4) ++ (-1.0, 0) coordinate (bar4);
\path (4) ++ (0, 0.4) node[invS] (4s) {\GraphTristateS} node[at = (4s.east), right]{Storage};

\path (4) ++ (0, 0) coordinate (4r);

\path[midArrow] (4s) -- node[below]{\pInPlot{3}} ++(-1.0, 0) coordinate (bar4s);

\path (4r) to ++(-1.0, 0) coordinate (bar4r);
\path[powerLine] (bar4s) ++(0, 0.1) node[yshift = 4] {\tiny $3$}  -- (bar4r) -- ++(0, -0.1);

\path[draw,  ->] (4r) ++(-1, 0) -- ++(0.2, 0.0) -- ++(0, -0.4)  node[right, yshift=5]{\wInPlot{3}};

% Vertical transmission lines
\path (bar1) ++ (0.0, -0.2) coordinate (l1bar1);
\path (l1bar1) ++ (0.2, 0.0) coordinate (l1bar1out);
\path (bar2) ++ (0.0, 0.2) coordinate (l1bar2);
\path (l1bar2) ++ (0.2, 0.0) coordinate (l1bar2out);
\path[midArrow] (l1bar1) -- (l1bar1out) -- node[right, rotate=90, anchor=north]{$p_{12}, q_{12}$} (l1bar2out) -- (l1bar2);

\path (bar3) ++ (0.0, 0.1) coordinate (l2bar3);
\path (l2bar3) ++ (-0.2, 0.0) coordinate (l2bar3out);
\path (bar4) ++ (0.0, 0.1) coordinate (l2bar4);
\path (l2bar4) ++ (-0.2, 0.0) coordinate (l2bar4out);
\path[midArrow] (l2bar4) -- (l2bar4out) -- node[left, rotate = 90, anchor=south]{$p_{34}, q_{34}$} (l2bar3out) -- (l2bar3);

% Horizontal transmission lines
\path (bar2) ++ (0.0, -0.2) coordinate (l3bar2);
\path (bar3) ++ (0.0, -0.1) coordinate (l4bar3);

\path[midArrow] (l3bar2) -- node[above]{$p_{24}, q_{24}$} (l4bar3);

% Diagonal transmission lines
\path(bar2) ++ (0.3, 0.0) coordinate (l5bar2out);
\path (bar4) ++ (0.0, 0.3) coordinate (l5bar4);
\path (l5bar4) ++ (-0.3, 0.0) coordinate (l5bar4out);
\path[midArrow] (bar2) -- (l5bar2out) -- node[below, xshift = 0, rotate=42.5]{$p_{23}, q_{23}$} (l5bar4out) -- (l5bar4);

\end{tikzpicture}
    \caption{Power system under investigation.}
    \label{fig:model}
\end{figure}

We consider the power system in Fig.~\ref*{fig:model}. 
It contains four nodes with a grid-forming unit connected to each one of them.
Conventional \ac{sg} units, are connected to nodes 1 and 4. 
Inverter-interfaced battery storage units are connected to nodes 2 and 3. 
Table~\ref*{tab:parameters_units} includes the parameters and power setpoints of the units. 
The power setpoints are chosen such that the conventional units provide power and the storage units are charging. 
The droop gains are chosen identical while the time constants of the filters are chosen such that the inverter-interfaced units exhibit slightly faster dynamics. 

\begin{table}[!t]
    \vspace{0.7em}
	\centering
    \caption{Unit parameters in per-unit system.}
	\begin{tabular}{rcccccc}
		\toprule
		 & $k_i^P$ &  $k_i^Q$ & $\tau_i^P$ & $\tau_i^Q$ & $p_i^d$ & $q_i^d$ \\
        \midrule
        Conventional units & $1\frac{1}{\text{pu}\, s}$ & $0.1$ & $1$s & $1$pu & $0.6$pu & $0$pu \\
        Storage units & $1\frac{1}{\text{pu}\, s}$ & $0.1$ & $0.3$s & $0.3$pu & $-0.25$pu & $0$pu \\
		\bottomrule
	\end{tabular}
	\label{tab:parameters_units}
\end{table}

The nodes are connected by a meshed grid with dominantly inductive lines and dominantly resistive shunts which include loads at nodes 1 and 3. 
The admittances of the lines and loads are shown in Table~\ref*{tab:parameters_grid}.
\begin{table}[!t]
	\centering
    \caption{Power system parameters}
	\begin{tabular}{rccc}
		\toprule
        & \makecell{Powerline\\$y_{ij}, \forall (i,j)\in \mathbb{E}$} & \makecell{Load\\$y_i^L,i\in\{1,3\}$} & \makecell{Shunt\\$y_i^S,\forall i \in \mathbb{V}$}\\
        \midrule
        Conductance $g$ & $2$pu & $0.38$pu & $0.02$pu\\
        Susceptance $b$ & $-20$pu & $-0.1$pu & $0.005$pu\\
		\bottomrule
	\end{tabular}
	\label{tab:parameters_grid}
\end{table}

\subsection{Data}
We simulated 1003 system trajectories over a time horizon of $50$s at a sampling time of $10$ms. 
They were obtained considering step changes of the voltage and frequency setpoints $v_i^d$ and $\omega_i^d$.
For each trajectory, we used 10 equidistantly spaced steps, which occured every $5$s. 
The magnitudes of the steps were sampled from uniform distributions with bounds $0.99$pu and $1.01$pu for $v_i^d$ as well as $2\pi \cdot 49.975$Hz and $2\pi \cdot 50.025$Hz for $\omega_i^d$. 
To train the \acp{node}, we split the data into four sets referred to as training-, validation-, test- and evaluation-dataset. 
The training-, validation- and test-dataset contain one trajectory each and were used for learning the parameters, validating improvement in the training loop and selecting the model with the best performance from the hyperparameter optimization, respectively. 
The evaluation-dataset contains the remaining 1000 trajectories and was used to assess the prediction accuracy of the best models. 

\subsection{Training, Validation and Evaluation}
\label{sec:training_validation_evaluation}
% We investigate the influence of different hyperparmeters on the performance of the models. 
We consider \acp{node} with three different solvers:  
two fixed-step integration methods, i.e., the Euler method and the \ac{rk4} method, and one variable-step integration method, i.e., the \ac{dp5} method. 
In what follows, we describe the procedure for obtaining the best model for each one of them.

We use the Adam optimizer~\cite{kingma_adam_2017} with constant learning rate $\alpha$ for adjusting the weights and biases of the \acp{node} by minimizing the objective~(\ref*{eq:loss_node}). 
Overall, we optimize the one-step-ahead (10ms) prediction over 5000 samples in the training-dataset. 
In each epoch, batch gradient descent is applied, i.e., we optimize over the entire training data at once. 
The number of epochs is limited to 10000 and early stopping is applied as soon as the objective, asserted on the validation data, does not decrease for 100 epochs. 
We choose the final weights and biases of the trained models, that first exhibited the lowest objective on the validation data.

\acp{nn} contain a large number of hyperparameters and choosing good ones manually can be cumbersome. 
Therefore, we made use of Bayesian optimization \cite{mockus_bayesian_1991}. 
Table \ref*{tab:ho_node} shows the hyperparameters for the \acp{node}, as well as the employed parameter ranges for the optimization.
\begin{table}[!t]
    \vspace{0.7em}
	\centering
    \caption{Hyperparameter search space for the NODEs.}
	\begin{tabular}{rc}
		\toprule
		Hyperparameter & Intervals/sets\\
        \midrule
        Hidden layers & $L \in [2,10]\subset \mathbb{N}$\\
		Hidden neurons/layer & $H \in [10,200] \subset \mathbb{N}$\\
        Learning rate & $\alpha \in [10^{-4},10^{-2}] \subset \mathbb{R}_{\geq 0}$\\
        Activation function & $h(\cdot) \in $\{ReLU, Sigmoid,Softplus\}\\
		\bottomrule
	\end{tabular}
	\label{tab:ho_node}
\end{table}
The hyperparameters are optimized over 150 trials, i.e., 150 fully trained models, in order to find good hyperparameters. 

For \ac{sindy}, we use the sequential-threshholded least-square algorithm~\cite{brunton_discovering_2016} for finding the coefficients of the candidate functions according to the objective~(\ref*{eq:loss_sindy}). 
We limit the number of iterations to 10000. 

For evaluating and comparing different \acp{node} and \ac{sindy}, we analyze the simulation \ac{rmse}
\begin{equation}
    \label{eq:rmse}
    \text{RMSE} = \sqrt{\frac{1}{K} \sum_{k=1}^{K} ||x({t_k}) - \hat{x}({t_k|t_0})||_2^2}
\end{equation}
where $x({t_k})$ is the true state of the reference system at time $t_k$ and $\hat{x}({t_k|t_0})$ is the prediction of the model at time $t_k$ given the initial condition $x(t_0)$.
Thus, we analyze the prediction accuracy of the models on each trajectory of the evaluation-dataset over a horizon of $50$s, i.e., $K = 5000$.

\subsection{Software and hardware}
All models and identification methods were implemented in Python. For \ac{sindy}, we used \texttt{pysindy} \cite{desilva2020}, for \acp{node} we used \texttt{pytorch} and \texttt{torchdiffeq} \cite{torchdiffeq}. Our \ac{node} implementation was inspired by \texttt{neuromancer} \cite{Neuromancer2023}. For hyperparemeter optimization we made use of \texttt{optuna} \cite{akiba2019optuna}. 

For training models, we used an Apple Mac mini 2023 with an M2 Pro (10-core CPU, 16-core GPU) and 16GB RAM.

\section{Results}
As described in Sec.~\ref{sec:experiment}, we compare \acp{node} with three different integration schemes, i.e. Euler, \ac{rk4} and \ac{dp5}, with each other and with \ac{sindy}. 
In what follows, we discuss the outcome of the hyperparameter optimization. 
Then, we will compare the prediction accuracy, using the best model of each \ac{node} with each other and with \ac{sindy}. 

\subsection{Influence of hyperparameters}
We discuss the choice of hyperparameters based on the one-step-ahead prediction accuracy on the test data, i.e., the basis on which the model was chosen during hyperparameter optimization.
Table~\ref{tab:ho_node_result} shows the hyperparameters for the \acp{node} resulting in the smallest \ac{mse}~\eqref{eq:loss_node} on the test data. 
For all three solvers, typically small networks performed better than larger ones: 
The Bayesian optimization ends in the lower bound of hidden layers. 
The optimal number of neurons lies close to the number of states in the model, i.e., $N_x = 11$. 
Apparently, small numbers are sufficient to capture the dynamics, however, more investigations with larger systems are needed to draw more general conclusions. 
Furthermore, $2 \cdot 10^{-3}$ to $5 \cdot 10^{-3}$ appears to be a good choice for the learning rate. 
For all models, using a continuously differentiable activation function, like the softplus function, is preferable over the ReLU function which is not differentiable. 
Even though continuous differentiability is technically required to backpropagate through the solver~\cite{kidger_neural_2022}, practically the ReLU function still performs better than the Sigmoid function. 
This might be due to vanishing gradient problems. 

Lastly, even though, the computational effort to train the \ac{node}, as indicated by training time per epoch, is significantly higher when a more sophisticated solver is used, the accuracy of the models only differs slightly: The \ac{mse}, which quantifies the quality of the one-step-ahead prediction on the test-dataset, is comparable for all three models, while the \ac{node} with \ac{dp5} performs a little better than the ones with Euler and \ac{rk4}.

\begin{table}[!t]
    \vspace{0.7em}
	\centering
    \caption{Hyperparameters for NODEs resulting in the smallest MSE on the test data.}
	\begin{tabular}{rccc}
		\toprule
	    Solver & Euler & \ac{rk4} & \ac{dp5} \\
        \midrule
		Hidden neurons & $12$ & $13$ & $12$\\
        Hidden layers & $2$ & $2$ & $2$ \\
        Learning rate & $3.99 \cdot 10^{-3}$& $2.15 \cdot 10^{-3}$ & $4.36 \cdot 10^{-3}$\\
        Activation function & Softplus & Softplus & Softplus \\
        \midrule
        Test data \ac{mse} (norm.)  & $1.01 \cdot 10^{-5}$& $1.07 \cdot 10^{-5}$ & $0.99 \cdot 10^{-5}$ \\
        Training time/epoch & $0.052$s & $0.076$s & $0.179$s\\
		\bottomrule
	\end{tabular}
	\label{tab:ho_node_result}
\end{table}

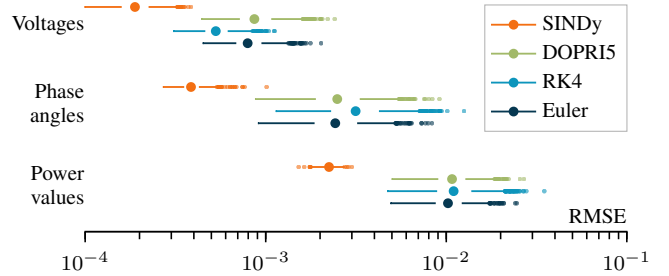
\begin{figure}
    \centering
    % !TeX encoding = UTF-8
% !TeX spellcheck = en_GB
% !TEX TS-program = pdflatex
% !TEX root = ../document.tex

\tikzset{external/export next=true}

\pgfplotsset{linestyle boxplot/.style={%
  boxplot = {%
    every box/.style={draw=none, fill=none},
    whisker extend=0,
    },
    mark=*,
    every mark/.append style={mark size=0.7pt, line width=0pt, opacity=0.6, fill=#1}, draw=#1,
    boxplot/draw/median/.code={%
          \draw[mark size=1.5pt, /pgfplots/boxplot/every median/.try]
          \pgfextra
          \pgftransformshift{
            \pgfplotsboxplotpointabbox
              {\pgfplotsboxplotvalue{median}}
              {0.5}
          }
          \pgfsetfillcolor{#1}
          \pgfuseplotmark{*}
          \endpgfextra
        ;
      },
  },
}

\begin{tikzpicture}[font=\footnotesize]
  \begin{semilogxaxis}[
    myPlot,
    clip=false,
    height = 45mm,
    width = 88mm,
    xmin = 1e-4,
    xmax = 1e-1,
    line width=0.7pt,
    xtick = {1e-6, 1e-5, 1e-4, 1e-3, 1e-2, 1e-1, 1e0, 1e1},
    y axis line style={white},
    ytick style={draw=none},
    yticklabels={},
    ymax = 11.5,
    clip=false,
    xlabel = {RMSE},
    x label style={at={(1, 0)}, anchor=south east, inner sep=0pt},
    legend style={at={(1, 1)}, anchor=north east, draw=black!30, thin, inner sep=2pt, legend cell align={left},},
    ]

  \addlegendimage{line legend, line width = 0.7pt, mark=*, mark size=1.5pt, color=vgOrange};
  \addlegendentry{SINDy};

  \addlegendimage{line legend, line width = 0.7pt, mark=*, mark size=1.5pt, color=vgGreen};
  \addlegendentry{DOPRI5};

  \addlegendimage{line legend, line width = 0.7pt, mark=*, mark size=1.5pt, color=vgLightBlue};
  \addlegendentry{RK4};

  \addlegendimage{line legend, line width = 0.7pt, mark=*, mark size=1.5pt, color=vgDarkBlue};
  \addlegendentry{Euler};

  \foreach \file in {datanew_p, datanew_theta, datanew_v}{
    \addplot [linestyle boxplot=vgDarkBlue, yshift=0.45em] table[y = fxu_euler_eval, col sep=comma]{data/\file.csv};
    \addplot [linestyle boxplot=vgLightBlue, yshift=0.15em] table[y = fxu_rk4adjoint_eval, col sep=comma]{data/\file.csv};
    \addplot [linestyle boxplot=vgGreen, yshift=-0.15em] table[y = fxu_dopri5_eval, col sep=comma]{data/\file.csv};
    \addplot [linestyle boxplot=vgOrange, yshift=-0.45em] table[y = sindy__eval, col sep=comma]{data/\file.csv};
  }

  \node[align=right, anchor=east, inner sep=0pt,] at (axis cs: 1e-4, 2.5) {Power \\ values};
  \node[align=right, anchor=east, inner sep=0pt,] at (axis cs: 1e-4, 6.5) {Phase \\ angles};
  \node[align=right, anchor=east, inner sep=0pt,] at (axis cs: 1e-4, 10.5) {Voltages};

  \end{semilogxaxis}

\end{tikzpicture}%
    \caption{\acp{rmse} with~\eqref{eq:rmse} obtained over 1000 datasets. The large dot displays the median. The white area around the dot marks the interquartile range, containing the middle $50\%$ of the values. The lines, so called whiskers, extend to each side to the values deviating up to $1.5$ times to the interquartile range from the bounds of the white area. The small dots are outliers.}
    \label{fig:prmsej_boxplots}
\end{figure}

\begin{figure}
    \vspace{0.4em}
    % !TeX encoding = UTF-8
% !TeX spellcheck = en_GB
% !TEX TS-program = pdflatex
% !TEX root = ../document.tex

\tikzset{external/export next=true}

\pgfplotsset{resultsPlot/.style={%
    clip = false,
    minor x tick num=1,
    grid=both,
    grid style={draw=black!25},
    major tick length=0pt,
    minor tick length=0pt,
    axis lines = left,
    axis line style= {-, draw opacity=0.0},
    y tick label style={
        /pgf/number format/.cd,
            scaled y ticks = false,
            fixed,
            precision=3,
        /tikz/.cd
        },
    height = 4cm,
		xtick = {0, 10, 20, 30, ..., 100},
		xmin = 0,
		xmax = 50,
		clip=false,
    width=8.5cm,
    legend columns=3,
    legend style={
      at={(0.5, 1.05)},
      anchor=south,
      draw=none,
      fill=none,
      legend cell align=left,
      /tikz/every even column/.append style={column sep=2.5mm},
      inner sep=0pt,
      legend cell align={left},
      },
  	}
}

\begin{tikzpicture}[font=\footnotesize]

\draw[draw=none, fill=none] (-1.45, 2.55) rectangle (7.2, -6.8);

\begin{axis}[%
    resultsPlot,
    xlabel = {},
    xticklabels = {~},
    ylabel={Voltage in pu},
    ymin = 0.988,
    ymax = 1.004,
    ytick={0.988, 0.992, ..., 1.008},
    minor y tick num=1,
]

% vgRed, vgOrange, vgYellow, vgGreen, vgLightBlue, vgDarkBlue

  \foreach \x [count=\i from 1] in {vgYellow, vgLightBlue, vgOrange, vgGreen}{
    \addplot[line width=0.7pt, color=black, forget plot] table [x=time, y=v\i, col sep=comma, each nth point=3, ]{data/trajectories_median_fxu_euler_0_1_2_3.csv};
  }

  \foreach \x [count=\i from 1] in {vgYellow, vgLightBlue, vgOrange, vgGreen}{
      \edef\temp{\noexpand
    \addplot[color=\x, line width=0.7pt, densely dashdotted] table [x=time, y=hatv\i, col sep=comma, each nth point=3, ]{data/trajectories_median_fxu_euler_0_1_2_3.csv};
      }
    \temp
  }

\addlegendentry{$\hat{v}_1$, $\hat{p}^{m}_{1}$}
\addlegendentry{$\hat{v}_2$, $\hat{\delta}_{12}$, $\hat{p}^{m}_{2}$}
\addlegendentry{$\hat{v}_3$, $\hat{\delta}_{13}$, $\hat{p}^{m}_{3}$}
\addlegendentry{$\hat{v}_4$, $\hat{\delta}_{14}$, $\hat{p}^{m}_{4}$}
\addlegendimage{line width=0.7pt, color=black}
\addlegendentry{True signal}

\end{axis}

\begin{axis}[%
  yshift = -3cm,
  resultsPlot,
  xticklabels = {~},
  xlabel = {},
  ylabel = {Angle diff. in rad},
  ymin = -0.09,
  ymax = 0.09,
  ytick={-0.12, -0.08, ..., 0.12},
  minor y tick num=1,
]

 \foreach \x [count=\i from 2] in {vgLightBlue, vgOrange, vgGreen}{
    \addplot[line width=0.7pt, color=black] table [x=time, y=theta1\i, col sep=comma, each nth point=3, ]{data/trajectories_median_fxu_euler_0_1_2_3.csv};
  }

  \foreach \x [count=\i from 2] in {vgLightBlue, vgOrange, vgGreen}{
      \edef\temp{\noexpand
    \addplot[color=\x, line width=0.7pt, densely dashdotted] table [x=time, y=hattheta1\i, col sep=comma, each nth point=3, ]{data/trajectories_median_fxu_euler_0_1_2_3.csv};
      }
    \temp
  }

\end{axis}

\begin{axis}[%
  yshift = -6cm,
  resultsPlot,
  ylabel={Power in pu},
  ymin = -0.5,
  ymax = 1,
  ytick={-2, -1.5, ..., 2},
  minor y tick num=1,
  xlabel = {Time in s},  % xticklabels = {},
  ]

  \foreach \x [count=\i from 1] in {vgYellow, vgLightBlue, vgOrange, vgGreen}{
    \addplot[line width=0.7pt, color=black] table [x=time, y=pm\i, col sep=comma, each nth point=3, ]{data/trajectories_median_fxu_euler_0_1_2_3.csv};
  }

  \foreach \x [count=\i from 1] in {vgYellow, vgLightBlue, vgOrange, vgGreen}{
      \edef\temp{\noexpand
    \addplot[color=\x, line width=0.7pt, densely dashdotted] table [x=time, y=hatpm\i, col sep=comma, each nth point=3, ]{data/trajectories_median_fxu_euler_0_1_2_3.csv};
      }
    \temp
  }

\end{axis}

\end{tikzpicture}%
    \caption{Predicted state of the system using the \ac{node} with Euler integration scheme for the trajectory that yielded the median \ac{rmse} with~\eqref{eq:rmse}.}
    \label{fig:simulation}
\end{figure}
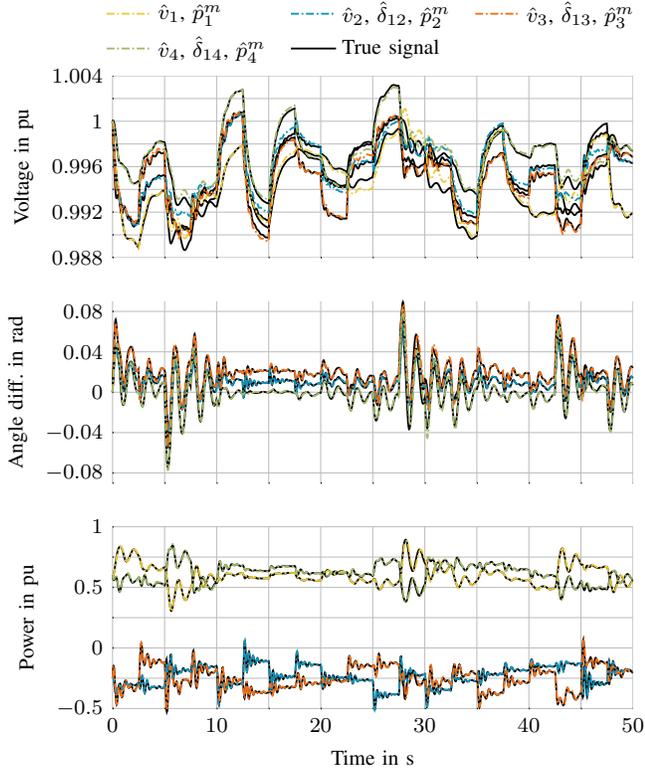

\subsection{Prediction accuracy}

Fig.~\ref{fig:simulation} shows the predicted state trajectories of all units using the \ac{node} with the Euler integration method for one of the trajectories of the evaluation dataset. 
While the predicted phase angle difference and power trajectories show good agreement with the true values, the predicted voltage trajectories show a slight deviation.
Similar behaviour can be observed for the other integration methods.

Fig.~\ref{fig:prmsej_boxplots} shows the boxplot of the simulation error over the entire prediction horizon for the four different models. 
The error is split into phase angles, power and voltages. 

\ac{sindy} clearly outperforms the \acp{node} in terms of prediction accuracy, yielding the lowest median \acp{rmse} on all states. 
Furthermore, the predictions of the model identified with \ac{sindy} exhibit comparibly low deviations from the median, displaying a robust performance over all datasets. 
It is important to note though, that we supplied the algorithm with a condensed set of candidate functions that are known to be part of the system dynamics which renders a comparison with \acp{node} almost unfair. 
In our experience, using a larger set of more trivial candidate functions did not result in a good \ac{sindy} model. 

Comparing \acp{node} with different integration schemes with each other, we see only small differences in the prediction accuracy. 
One would expected the \ac{node} using \ac{dp5}, to outperform the \acp{node}, using fixed-step solvers, especially Euler. 
However, in practice, the \ac{node} with \ac{dp5} performs slightly worse than the Euler method in terms of phase angle, voltage and power predictions. 
A possible cause for this observation is discussed in \cite{ott_resnet_2023}: The learned model is closely tied to the used integration method and stepsize, which is a result from training the network with respect to its interaction with the solver. 
The author of \cite{kidger_neural_2022} refers to this as \enquote{baked-in discretization}. 
As long as the sampling time of the data does not change, this can be beneficial as we obtain an accurate model with a light-weight solver, rendering training, as well as inference, more computationally efficient. 
When learning models from irregularily sampled data, \acp{node} with a variable step solver, like \ac{dp5}, should be preferred as the model is then not tied to a specific step size. However, this comes with significantly more computational effort in training and simulation. 

\section{Conclusion and Outlook}
In this paper, we investigated the capabilities of \acp{node} to capture the dynamics of coupled droop-controlled grid-forming units. 
We compared the performance of \acp{node} with different solvers to \ac{sindy} in terms of prediction accuracy and computing times per epoch. 
We concluded that \ac{sindy} outperforms the \acp{node} significantly, but only when choosing a very specific set of candidate functions. 
Meanwhile, the choice of solver used with the \acp{node} does not result in a significant difference in the prediction accuracy, even though the training effort increases when using higher-order solvers. 
From the hyperparameter optimization, we conclude that small networks with non-saturating, continuously differentiable activations functions, like softplus, appear preferable over larger ones and ones with a activation functions that do not exhibit the formerly stated characteristics. 

In the future, we intend to investigate more complex grid topologies and dynamics. 
Specifically, we want to find out if the number of neurons in the hidden layers should be chosen close to the number of states of the system. 
Furthermore, we will consider different data. 
Specifically, we aim to use power factors instead of phase angle measurements, and incorporate noise. 
Lastly, dealing with missing data from certain states will be investigated. 

\bibliographystyle{template/IEEEtran}
\bibliography{template/IEEEabrv,IEEEexample}

\end{document}